\RequirePackage{fix-cm}
\documentclass[twocolumn,epjc3]{svjour3}  
\smartqed  
\RequirePackage{graphicx}
\journalname{Eur. Phys. J. C}
\begin{document}

\title{Low-redshift estimates of the absolute scale of baryon acoustic oscillations
}


\author{Thais Lemos\thanksref{e1,addr1}
        \and
        Ruchika\thanksref{e2,addr2} 
        \and
        Joel C. Carvalho\thanksref{e3,addr1}
        \and
        Jailson Alcaniz\thanksref{e4,addr1}
}

\thankstext{e1}{e-mail: thaislemos@on.br}
\thankstext{e2}{e-mail: ruchika.ruchika@roma1.infn.it}
\thankstext{e3}{e-mail: jcarvalho@on.br}
\thankstext{e4}{e-mail: alcaniz@on.br}


\institute{Observat\'orio Nacional - Rio de Janeiro - RJ, 20921-400, Brazil \label{addr1}
           \and
           INFN Sezione di Roma, Piazzale Aldo Moro 2, I-00185 Rome, Italy\label{addr2}
}

\date{Received: date / Accepted: date}

\maketitle

\begin{abstract}
Measurements of the characteristic length scale $r_s$ of the baryon acoustic oscillations (BAO) provide a robust determination of the distance-redshift relation. Currently, the best (sub-per cent) estimate of $r_s$ at the drag epoch is provided by Cosmic Microwave Background (CMB) observations assuming the validity of the standard $\Lambda$CDM model at $z \sim 1000$. Therefore, inferring $r_s$ from low-$z$ observations in a model-independent way and comparing its value with CMB estimates provides a consistency test of the standard cosmology and its assumptions at high-$z$. 
In this paper, we address this question and estimate the absolute BAO scale combining angular BAO measurements and type Ia Supernovae data. Our analysis uses two different methods to connect these data sets and finds a good agreement between the low-$z$ estimates of $r_{s}$ with the CMB sound horizon at drag epoch, regardless of the value of the Hubble constant $H_0$ considered. These results highlight the robustness of the standard cosmology at the same time that they also reinforce the need for more precise cosmological observations at low-$z$.

\end{abstract}

\section{Introduction}\label{sec1}

The standard $\Lambda$-Cold Dark Matter ($\Lambda$CDM) model successfully describes the current cosmological observations. Among these observations, measurements of the Cosmic Microwave Background (CMB) and Type Ia Supernovae (SNe Ia) have been used to infer the angular diameter distance $d_{A}(z)$ out to $z \approx 1100$ and luminosity distances $d_{L}(z)$ out to $z \approx 2$, respectively, being also highly complementary tools for measuring the cosmic expansion history and constraining cosmological parameters such as the matter density parameter, $\Omega_{m}$, and the local expansion rate, $H_{0}$ (see e.g., \cite{Weinberg:2013agg} and references therein)\footnote{See also \cite{DiValentino:2021izs} for a recent review on possible tensions involving estimates of the $\Lambda$CDM model parameters.}.  

On the other hand, the position of the baryon acoustic oscillation (BAO) feature observed in the large-scale distribution of galaxies is determined by the comoving sound horizon size at the drag epoch,
\begin{equation}
r_{s} (z_{\rm{drag}})=r_{d} = \int_{z_{\rm{drag}}}^{\infty}{\frac{c_s(z)}{H(z)}}dz\;,
\end{equation}
where $c_{s}(z)$ is the sound speed of photon-baryon fluid and $z_{\rm{drag}} \approx 1100$ is the redshift at which baryons were released from photons. Such a  characteristic scale provides a fundamental standard ruler that can be measured in the CMB anisotropy spectrum and distribution of large-scale structure at lower $z$, and used to estimate cosmological parameters \cite{aghanim}. However, as well known, the comoving length $r_{d}$ is calibrated at $z \sim 1000$ using a combination of observations and theory, which makes its estimates vulnerable to systematic errors from possible unknown physics in the early universe \cite{Sutherland}.

It is worth mentioning that although the BAO feature evolves by a small amount during cosmic evolution \cite{Padmanabhan:2012hf}, it is undoubtedly the most robust cosmic ruler at intermediate redshifts currently available. Moreover, its length scale also plays a role in the discussions about the current tensions in the standard cosmology, as some possible solutions for the mismatch between local measurements of $H_{0}$ and the value inferred from CMB observations assuming the $\Lambda$CDM model, known as the Hubble tension, suggest an increase of the pre-recombination expansion rate, which implies a reduction of the sound horizon at recombination with an increase in $H_{0}$ (see, e.g. \cite{Karwal:2016vyq,bernal,Alcaniz:2019kah,Kamionkowski:2022pkx} and references therein). More importantly, independent estimates of $r_{s}$ can be used as a probe of the standard assumptions of the early universe cosmology.

In this paper, we address this latter issue and derive estimates of the absolute BAO scale (hereafter denoted as $r_{s}$) using only low-$z$ measurements in a model-independent way. In order to perform our analysis, we use two methods to combine SNe measurements from the Pantheon compilation \cite{Scolnic} with eleven angular BAO measurements derived from public data of the Sloan Digital Sky Survey (SDSS) and reported in \cite{Carvalho2016,Carvalho2017}. We compare our estimates of $r_{s}$ with the sound horizon at drag epoch $r_{d}$ and discuss potential mismatches between them, considering different measurements and estimates of the Hubble constant $H_{0}$. 

We organize this paper as follows. In Section 2, we briefly review the physics of the BAOs and the need for independent estimates of $r_{s}$. We describe the data used in our analysis and the methodology proposed in Sections 3 and 4, respectively. Our results are discussed in Section 4. In Section 5, we present our main conclusions.

\begin{table*}
\centering
\caption{2D BAO measurements from angular separation of pairs of galaxies.}
\begin{tabular}{| c | c | c | c | c || c | c | c |}
\hline
$z_{\rm{bin}}$ & $\Delta z_{\rm{bin}}$ &$\theta_{BAO}(z)[^{\circ}]$ &  Reference & $z_{\rm{bin}}$ & $\Delta z_{\rm{bin}}$ & $\theta_{BAO}(z)[^{\circ}]$ & Reference\\
\hline
0.45 & [0.44, 0.46] & 4.77 $\pm$ 0.17 & \cite{Carvalho2016} & 0.57 & [0.56, 0.58] & 4.59 $\pm$ 0.36 & \cite{Carvalho2017} \\
0.47 & [0.46, 0.48] & 5.02 $\pm$ 0.25 & \cite{Carvalho2016} & 0.59 & [0.58, 0.60]& 4.39 $\pm$ 0.33 & \cite{Carvalho2017} \\
0.49 & [0.48, 0.50] & 4.99 $\pm$ 0.21 & \cite{Carvalho2016} & 0.61 & [0.60, 0.62]& 3.85 $\pm$ 0.31 & \cite{Carvalho2017} \\
0.51 & [0.50, 0.52]& 4.81 $\pm$ 0.17 & \cite{Carvalho2016} & 0.63 & [0.62,0.64] & 3.90 $\pm$ 0.43 & \cite{Carvalho2017} \\
0.53 & [0.52, 0.54]& 4.29 $\pm$ 0.30 & \cite{Carvalho2016} & 0.65 & [0.64, 0.66]& 3.55 $\pm$ 0.16 & \cite{Carvalho2017} \\
0.55 & [0.56, 0.57]& 4.25 $\pm$ 0.25 & \cite{Carvalho2016} &  &  & & \\
\hline
\end{tabular}
\label{tab:BAO_data}
\end{table*}

\section{BAO features}\label{sec2} 

Baryon Acoustic Oscillations arise due to the competing effects of radiation pressure and gravity in the early Universe. When photons and baryons decoupled, the sound waves freeze out, leaving a fundamental scale in large-scale structure in the Universe \cite{Peebles,Sunyaev}. Since such
scale remains imprinted in the galaxy distribution, BAO can be considered a standard ruler which, combined with CMB and SNe measurements, places the best constraints on cosmological parameters today, including on the dark energy equation-of-state parameter and the spatial curvature \cite{Weinberg:2013agg}.

The BAO feature can be separated into transversal and radial modes, providing independent estimates of angular distance diameter and the expansion rate $H(z)$ \cite{Seo2003,Aubourg:2014yra}. Most of the current BAO measurements constrain the quantity $r_s/D_V$, where $D_V$ is  the dilation scale defined as \cite{SDSS:2005xqv}
\begin{equation}
D_V(z) =  \left[(1+z)^2 {d}^2_{A}(z) \frac{cz}{H(z)}\right]^{1/3}.   
\end{equation}
In this 3D approach, the BAO scale is obtained by applying the spatial 2-point correlation function to a large
distribution of galaxies and assuming a fiducial cosmology to transform the measured angular positions and redshifts into comoving distances. On the other hand, it is possible to obtain fully model-independent BAO measurements from the angular 2-point correlation function (2PACF) - here referred to as 2D BAO, which involves only the angular separation $\theta$ between pairs of galaxies. Using thin-enough redshift bins, one measures the angular BAO scale given by
\begin{equation}\label{eq2}
\theta_{\rm{BAO}}(z)=\frac{{r}_{s}}{(1+z){d}_{A}(z)} \;.
\end{equation}

As mentioned earlier, there is a slight difference between the comoving sound horizon at the drag epoch $r_d$ and the comoving length scale of the BAO feature in a galaxy survey $r_s$, which in principle results from the non-linear growth of structure and evolution of perturbations \cite{Padmanabhan:2012hf}. Currently, the best-inferred value of the sound horizon, $r_{d} = 147.21 \pm 0.23$ Mpc, was determined from the CMB power spectrum by the Planck mission \cite{Planck2018}. Such an estimate is obtained in the context of the $\Lambda$CDM model and does not consider possible new or unknown physics at earlier times\footnote{The above estimate of $r_d$ is obtained assuming the standard recombination history with the effective number of neutrino species $N_{eff} = 3.046$ and the usual evolution of matter and radiation energy densities.}. Therefore, measuring $r_s$ at low-$z$ and comparing it with the sound horizon estimates from CMB constitutes an important consistency test of the $\Lambda$CDM model and its assumptions at $z \sim 1000$. This idea was first discussed by \cite{Sutherland} who derived an accurate approximation relating the BAO
dilation scale $D_V (z)$ to the luminosity distance $d_L(z)$ that can be used to determine the length of the horizon scale at low-redshifts. In what follows, we closely follow this idea using a different method and data sets and estimate the absolute BAO scale in a model-independent way by considering current measurements of the angular BAO scale in combination with SNe data.

\section{Data sets}\label{sec3} 

To estimate the absolute scale of baryon acoustic oscillations $r_{s}$, we use a set of 11 $\theta_{\rm{BAO}}(z)$ measurements obtained from public data of the Sloan Digital Sky Survey (SDSS), namely DR10, DR11, and DR12 \cite{Carvalho2016,Carvalho2017}. As mentioned earlier, these measurements are derived by calculating the 2PACF between pairs of objects  and considering thin redshift slices with a fair number of cosmic tracers. It is worth mentioning that in such an approach, the measurement errors are determined by the width of the BAO bump, which in general leads to larger error bars when compared with the 3D approach.  The compiled 2D BAO dataset is shown in Table \ref{tab:BAO_data} (we refer the reader to \cite{Carvalho2016,sanchez,Menote:2021jaq} for a detailed discussion on these measurements).

We also use the Pantheon Sample \cite{Scolnic}, which comprises 1048 SNe data points ranging in the redshift interval $0.01 \leq  z \leq  2.3$. This compilation includes 279 SNe ($0.03 \leq z \leq  0.68$) discovered by Pan-STARSS1 (PS1) Medium Deep Survey with distance estimates from SDSS, SNLS, and various low-$z$ SNe along with HST samples. These data have been corrected for bias corrections in the light curve fit parameters using the BEAMS with Bias Corrections (BBC) method. Therefore, the systematic uncertainty related to the photometric calibration has been substantially reduced. Corrected magnitudes of the 1048 SNe, along with their redshift, can be found in \cite{Scolnic}.


\begin{table*}[h]
\centering
\caption{Estimates of the absolute BAO scale from 2D BAO and SNe data for binning and GP methods. In this analysis, we assume the value of absolute magnitude as $M_{B}$=-19.214 $\pm$ 0.037 (SH0ES 2021$_{a}$) \cite{Efstathiou}.}
\scalebox{0.65}{
\renewcommand{\arraystretch}{1.3}
\begin{tabular}{|c|c|c|c|c|c|c|c|c|c|c|c|c|}
\hline
2D BAO &  & & Type Ia SNe & Binning  &  &  & & &GP & & & \\ 
\hline
\hline
$n$-th bin &  $z_{\rm{bao}}$ & $\theta_{\textrm{BAO}}$& $[z^{\rm a}_l,z^{\rm a}_r]$ & ${z}_{\rm{SN}}$ & $n_a$ & $d_{L}$ (Mpc) & $d_{A}$ (Mpc) & $r_{s}$(Mpc) & $z_{\rm{GP}} = z_{\rm{bao}}$ & $d_{L}$ (Mpc) & $d_{A}$ (Mpc) & $r_{s}$(Mpc) \\ \hline
$1$ &  $0.45$ & 4.77 $\pm$ 0.17 & [0.44007, 0.45173]  & 0.44730 & 10 & 2304.34 $\pm$ 226.62 & 1096.00 $\pm$ 107.79 & 132.39 $\pm$ 13.85 &0.45 & 2361.35 $\pm$ 45.00 & 1123.06 $\pm$ 21.40 & 135.57 $\pm$ 5.48\\ \hline
$2$ & $0.47$ & 5.02 $\pm$ 0.25 & $[0.4664, 0.47175]$  & 0.46925 & 7 & 2606.23 $\pm$ 300.43 & 1206.08 $\pm$ 139.03 & 155.20 $\pm$ 19.49 & 0.47& 2486.40 $\pm$ 47.71 & 1150.42 $\pm$ 22.07  & 148.18 $\pm$ 7.91\\\hline
$3$ & $0.49$ & 4.99 $\pm$ 0.21 & [0.4804, 0.49737]   & 0.48635 & 7 & 2581.83 $\pm$ 206.25 & 1162.93 $\pm$ 92.90 & 150.69 $\pm$ 13.61  & 0.49 & 2609.44 $\pm$ 50.48 & 1175.31 $\pm$ 22.74 & 152.52 $\pm$ 7.06 \\ \hline
$4$ & $0.51$ & 4.81 $\pm$ 0.17 & [0.50718, 0.51476]  & 0.51092 & 9 & 2753.04 $\pm$ 246.42 & 1207.42 $\pm$ 108.07 & 153.13 $\pm$ 14.74  & 0.51 & 2734.46 $\pm$ 53.36 & 1199.05 $\pm$ 23.40 & 152.01 $\pm$ 6.14\\ \hline
$5$ & $0.53$ &  4.29 $\pm$ 0.30 & [0.52851, 0.53433]  & 0.53235 & 4 & 2697.60 $\pm$ 203.15 & 1152.38 $\pm$ 86.78 & 132.04 $\pm$ 13.57   & 0.53 & 2861.86 $\pm$ 56.19 & 1222.48 $\pm$ 24.00  & 140.05 $\pm$ 10.17 \\ \hline
$6$ & $0.55$ & 4.25 $\pm$ 0.25 & [0.54539, 0.55381]  & 0.55009 & 6 & 2958.21$\pm$ 286.74 & 1231.30 $\pm$ 119.35 & 141.56 $\pm$ 16.05   & 0.55 & 2996.40 $\pm$ 58.93 & 1246.98 $\pm$ 24.52  & 143.38 $\pm$ 8.89 \\ \hline
$7$ & $0.57$ & 4.59 $\pm$ 0.36 & [0.565, 0.575]  & - & 0 &  - & - & -  & 0.57 & 3134.79 $\pm$ 61.61 & 1271.70 $\pm$ 24.99 & 159.95 $\pm$ 12.93\\ \hline
$8$ & $0.59$ & 4.39 $\pm$ 0.33 & [0.58575, 0.59185]  & 0.58878 & 5  & 3304.57 $\pm$ 286.21 & 1307.13 $\pm$ 113.21 & 159.31 $\pm$ 18.27  & 0.59 & 3275.58 $\pm$  64.78 & 1295.44 $\pm$ 25.62 & 157.83 $\pm$ 12.27\\ \hline
$9$ & $0.61$ & 3.85 $\pm$ 0.31 & [0.60825, 0.61124]  & 0.61 & 4 & 3551.85 $\pm$ 328.27 & 1370.26 $\pm$ 126.64 & 148.43 $\pm$ 18.05  & 0.61 & 3409.94 $\pm$  68.98 & 1315.44 $\pm$ 26.61 & 142.31 $\pm$ 11.81 \\ \hline
$10$ & $0.63$ & 3.90 $\pm$ 0.43 & [0.62522, 0.63222]  & 0.62964 & 6 & 3346.38 $\pm$ 340.44 & 1259.50 $\pm$ 128.13 & 140.10 $\pm$ 20.99  & 0.63 & 3536.01 $\pm$ 74.38 & 1330.64 $\pm$ 27.99 & 147.65 $\pm$ 16.57\\ \hline
$11$ & $0.65$ & 3.55 $\pm$ 0.16  & [0.64191, 0.64864]  & 0.64509 & 5 & 3654.40 $\pm$ 307.19 & 1342.29 $\pm$ 112.83 & 137.20 $\pm$ 13.09   & 0.65 & 3651.93 $\pm$ 79.92 &1341.31 $\pm$ 29.35 & 137.13 $\pm$ 6.87 \\ \hline
\end{tabular}
}
\label{tab:BAO_bingp}
\end{table*}

\begin{table*}[h]
\centering
\caption{Estimates of the absolute BAO scale for different values of $H_{0}$. The parameter $\eta$ is defined as the ratio $\frac{r_{s}}{r_{d}}$, whereas the parameter $\sigma$ quantifies the 
difference between $r_s$ and CMB inferred sound horizon at drag epoch, $r_{d}= 147.09 \pm 0.26$ Mpc.}
\scalebox{0.9}{
\renewcommand{\arraystretch}{1.3}
\begin{tabular}{|c|c|c|c|c|c|c|c|c}
\hline
\multicolumn{2}{|c|}{ } &  \multicolumn{3}{|c|}{ Binning } &  \multicolumn{3}{|c|}{ G.P.}\\
\hline
\hline
Measurement & $H_{0}$(Km/s/Mpc) & $r_{s}$(Mpc)   & $\eta = \frac{r_s}{r_d}$ & $\sigma$ & $r_{s}$(Mpc) & $\eta$ & $\sigma$\\ 
\hline
\hline
Planck CMB + Lensing & 67.36 $\pm$ 0.54   & 159.44 $\pm$ 17.88    & 1.08 & 0.7 & 161.59 $\pm$ 10.96 & 1.1 & 1.32\\ \hline
ACT + WMAP CMB &  67.6 $\pm$ 1.1 & 158.70 $\pm$ 17.93  & 1.08 & 0.65  & 160.84 $\pm$ 11.13 & 1.09 & 1.23\\ \hline
BOSS DR12 + BBN & 68.5 $\pm$ 2.2  & 156.53 $\pm$ 18.21   & 1.06 &  0.52 & 158.64 $\pm$ 11.83 & 1.08 & 0.98\\ \hline
SH0ES 2021 & 73.2 $\pm$ 1.3  & 146.82 $\pm$ 16.76    & 0.99 & 0.02 & 148.80 $\pm$ 10.35 & 1.01 & 0.16\\ \hline
Masers &  73.9 $\pm$ 3.0 & 145.41 $\pm$ 17.29   & 0.99 & 0.10 & 147.37 $\pm$ 11.57 & 1.00 & 0.02\\ \hline
SH0ES 2019 &  74.0 $\pm$ 1.4 & 145.14 $\pm$ 16.44   & 0.99 & 0.12 & 147.10 $\pm$ 10.25 & 1.00 & 0.00\\ \hline
SH0ES 2021$_{a}$ &  74.1 $\pm$ 1.3  & 145.01 $\pm$ 16.39    & 0.98 & 0.13 & 146.96 $\pm$ 10.19 & 0.99 & 0.013\\ \hline
Tully Fisher & 76.0 $\pm$ 2.6  & 141.45 $\pm$ 16.45   & 0.96 & 0.34  &  143.35 $\pm$ 10.80 & 0.97 & 0.35\\ \hline
\hline
\end{tabular}
}
\label{tab:all_results}
\end{table*}

\section{Analysis and Results}

It is possible to estimate the BAO scale $r_{s}$ from low-$z$ observations in a model-independent way using Eq. (\ref{eq2}) if one knows {\it{i)}} transversal BAO measurements and {\it{ii)}} angular diameter distances, both at the same redshift without assuming a fiducial cosmology. We met the first requirement by using the $\theta_{\rm{BAO}}(z)$ values displayed in Table \ref{tab:BAO_data}. For completing the second requirement, we use the SNe data discussed above and convert the SNe distance modulus ($\mu_{0}(z) = m_{B}^0(z) - M_{B}$) into luminosity distances using
\begin{equation}\label{dl}
    d_{L}(z)=10^{(\mu_{0}(z) - 25)/5}\;,
\end{equation}
where $m_{b}(z)$ and $M_{B}$ are the apparent and absolute magnitude of SNe, respectively. Initially we assume $M_{B} = - 19.214 \pm 0.037$ mag, as obtained by the SH0ES collaboration combining geometrical distance estimates from Detached Eclipsing Binaries in LMC \cite{piet}, MASER NGC4258 \cite{reid}, and recent parallax measurements of 75 Milky Way Cepheids with HST photometry \cite{riess21} and GAIA Early Data Release 3 (EDR3, \cite{lind20a}). We then use the standard distance-duality relation $d_{L}(z) = (1 + z)^2d_{A}(z)$ to obtain a set of theory-independent estimates of angular diameter distances\footnote{In recent years, several analyses have observationally tested the distance-duality relation and verified its validity with a good precision (see, e.g., \cite{Holanda:2012at,Ellis:2013cu,Goncalves:2019xtc}).}.

\begin{figure*}
\begin{center}
\includegraphics[width=0.455\textwidth]{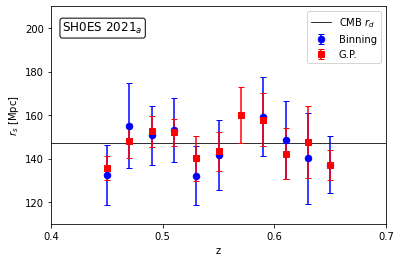}
\hspace{1.0cm}
\includegraphics[width=0.45\textwidth]{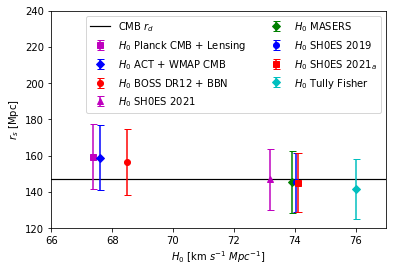}
\caption{{\textit{Left)}} The acoustic scale of BAO from transversal (2D) BAO and SNe datasets assuming $H_0 = 74.1 \pm 1.3$ $\rm{km.s^{-1}.Mpc^{-1}}$. The red points are obtained from Gaussian Processes, while the blue points are obtained from binning the Pantheon dataset. The black horizontal line in all figures denotes the $\Lambda$CDM estimate from CMB data ($r_{d}$). {\textit{Right)}} Estimates of the acoustic scale of BAO considering the different values of $H_{0}$ discussed in the text.}
\label{fig:res2021}
\end{center}
\end{figure*}

In order to obtain $d_{A}$ values at approximately the same redshifts of the $\theta_{\rm{BAO}}(z)$ measurements, we adopt two methods:

\begin{itemize}
\item {\it{Binning SNe Sample:}} We consider SNe within the redshift interval $0.44 \leq z \leq 0.66$ and group the data into 12 redshift bins centered at a $z_{\rm{SN}}$, which corresponds to the mean of all SNe redshifts inside the BAO bin interval $z_{\rm{bin}}$ (see Table \ref{tab:BAO_data}). However, we find no SNe in the bin interval $\Delta z_{\rm{bin}} = 0.565 - 0.575$, which explains why we do not show binning results for the BAO measurement at $z_{\rm{BAO}} = 0.57$ in Table \ref{tab:BAO_bingp} and Figure \ref{fig:res2021}.

\item {\it{Gaussian Process:}} We also apply the Gaussian Process (GP) method to reconstruct the SNe data. We use
the GaPP python library (for details of GaPP\footnote{https://github.com/carlosandrepaes/GaPP}, see \cite{GaPP}) with a square exponential covariance function and optimize its hyperparameters by maximizing the GP’s likelihood to obtain the reconstruction $m(z)$ and derive $d_{A}(z)$ at the same redshift of the $\theta_{\rm{BAO}}(z)$ measurements (we refer the reader to \cite{Shafieloo:2012ht,Li:2015nta,Gonzalez:2016lur,OColgain:2021pyh,Briffa:2020qli} and references therein for detailed discussions of GP reconstructions). 
\end{itemize}

From the values of $\theta_{\rm{BAO}}(z)$ and $d_{A}(z)$, we can estimate the absolute BAO scale, $r_{s}$. To calculate its uncertainty $\sigma_{r_{s}}$, we consider the errors associated with the binning process and SNe observations for the binning method, whereas, for the GP reconstruction, only the errors in the SNe apparent magnitude are considered. The results obtained from both methods are shown in Table \ref{tab:BAO_bingp}. It is important to observe that they agree within $1\sigma$ level for all bins, as also shown in Fig. \ref{fig:res2021} (left).

Using one particular value of absolute magnitude $M_{B}$ makes our results depend on the choice of $M_{B}$. Thus, we also derive the BAO scale $r_{s}$ by considering other values of $M_{B}$ or, equivalently, of $H_{0}$. For that, we use Eq. (9) of \cite{Riess:2016jrr}
\begin{equation}\label{mbh0}
M_{B} = 5(\log H_{0} - \alpha_B - 5)\;,
\end{equation}
where $\alpha_{B} = 0.71273 \pm 0.00176$ is the intercept of the Hubble diagram obtained  model independently from low-$z$ SNe ($0.023 \leq z \leq 0.15$). In what follows, we consider eight Hubble constant measurements, including the values obtained by the Planck CMB+Lensing \cite{aghanim}, ACT+WMAP CMB \cite{aiola}, and SH0ES \cite{Reiss2019,Reiss2021} collaborations and values derived from BOSS DR12+BBN \cite{Damico}, Masers \cite{Masers}, and the Tully-Fisher relation \cite{TFR}. We also use the $H_0$ value from a recent SN study \cite{Efstathiou} (denoted as SH0ES 2021$_{a}$), slightly higher than SH0ES 2021 values due to different period ranges and photometric samples. We use Eq. (\ref{mbh0}) to calculate the corresponding $M_{B}$, as described earlier. 

\begin{figure}
\includegraphics[width=0.48\textwidth]{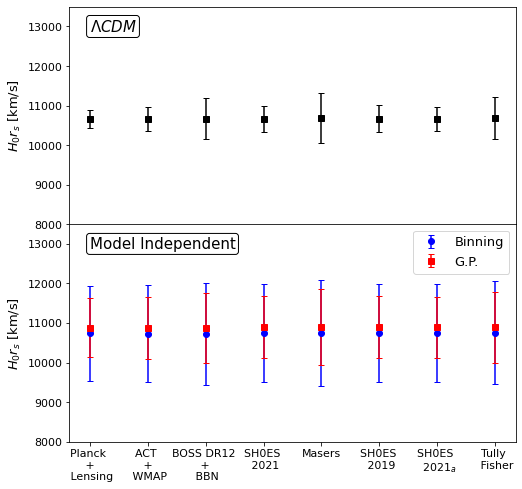}
\caption{The product of $H_{0} \times r_{s}$ for all low and high redshift estimates of $H_{0}$  fixing the $\Lambda$CDM model (upper panel) and performing the model-independent analysis (lower panel) discussed in the text. }
\label{fig:H0rs}
\end{figure}

In Table \ref{tab:all_results} and Fig. \ref{fig:res2021} (right), we show the average values of the 11 estimates of $r_{s}$ derived for the eight $H_{0}$ measurements considered in the analysis. The corresponding figures of each $H_0$ measurement separately are given in the \ref{Appendix}. The results for both methods show that low-$z$ measurements, which prefer higher values of the Hubble constant, support smaller $r_{s}$ values than PLANCK, WMAP, and BBN high-$z$ estimates. This result aligns entirely with the fact that the product of the Hubble constant and BAO acoustic scale is constant, as seen in Fig. \ref{fig:H0rs} (see also \cite{evslin}). Whereas the lower panel of that figure shows that both the binning and GP constraints on this product coincide within $1\sigma$, in the top panel, we see that fixing the $\Lambda$CDM model, the error bars decrease significantly. It is worth mentioning that, differently from the $\Lambda$CDM estimate obtained from CMB data, which furnishes a sub-percent estimate of the sound horizon at drag epoch from CMB observations ($r_{d}= 147.09 \pm 0.26$ Mpc), our model-independent approach combining current BAO and SNe data provides  $\sim 10\%$-error estimates. Such errors come mainly from the $\theta_{\rm{BAO}}(z)$ measurements and, because of this current uncertainty, the mean values of $r_{s}$ for both low and high-$z$ measurements show a good agreement (within $1\sigma$) with the standard model estimate of $r_{d}$. 

For completeness, we also estimate the absolute scale of BAO using low-$z$ measurements of the ratio $D_{V}/r_{s}$. For that, we use the two BAO measurements at $z = 0.106$ \cite{beutler} and $z = 0.150$ \cite{ross} and consider the expression derived by \cite{Sutherland}
\begin{equation}\label{eq:suth}
    D_{V}(z) \simeq \frac{3}{4} d_{L} (\frac{4}{3}z )(1+\frac{4}{3}z )^{-1}(1-0.0245z^{3}+0.0105z^{4}),
\end{equation}
which provides a quite accurately approximation till redshift $z \le 0.4 $ for nearly all flat and accelerating models. 
The results are shown in Table \ref{tabledv}.  For both 
GP and Binning methods, they show good agreement with the sound horizon estimate obtained by CMB observations assuming the $\Lambda$CDM model and attest to the robustness of the standard model assumptions at high-$z$.

\begin{table}
\label{dv2}
\caption{Estimates of $r_{s}$ from the combination of $D_{V}$ and SNe data.}
\scalebox{0.865}{
\begin{tabular}{| c | c | c |c | c |}
\hline
 & Binning & &  & \\
\hline
${z}$ & $d_{L}({4/3z})$ & $D_{V}({z})$ & $r_{s}({z})(Mpc)$ & $\sigma$ \\
\hline
0.106 & 652.42 $\pm$ 36.37 &  428.67 $\pm$ 23.89 & 144.03 $\pm$ 10.29 & 0.30\\
0.150 & 935.72 $\pm$ 59.54 & 584.66 $\pm$ 37.21 & 130.84 $\pm$ 35.33 & 0.46\\
\hline
\hline
 &  G.P. & & & \\
 \hline
 ${z}$ & $d_{L}({4/3z})$ & $D_{V}({z})$ & $r_{s}({z})(Mpc)$ & $\sigma$\\
 \hline
 0.106 & 652.29 $\pm$ 39.73  &  428.60 $\pm$ 26.10 &  144.01 $\pm$ 10.88 & 0.28\\
0.150 &  979.42 $\pm$ 72.48 & 611.74 $\pm$ 45.27  & 136.90 $\pm$ 37.78 & 0.27\\
\hline
\end{tabular}
}
\label{tabledv}
\end{table}

\section{Conclusions}\label{sec6}

The standard cosmology has faced several tensions with observational data and their increased accuracy in recent years. The most significant is the $\approx 5\sigma$ discrepancy between the values of the Hubble constant obtained from distance measurements of galaxies in the local universe calibrated by Cepheids and low-$z$ SNe and the CMB estimate assuming the standard $\Lambda$CDM model. These observational discrepancies, as well as the lack of a satisfactory theoretical description of the dark energy, motivate the need to probe the consistency of the model.

In this paper, we tested the consistency of the $\Lambda$CDM assumptions at $z \sim 1000$ by estimating the absolute BAO scale, $r_{s}$, from low-$z$ observations and comparing it with the sound horizon estimate obtained from current CMB observations. As well known, the latter is derived by assuming General Relativity, the standard recombination history with the effective number of neutrino species $N_{eff} = 3.046$ and the usual evolution of matter and radiation energy densities, and predicts $r_{d} \approx r_{s}$. Models that violate at least one of these assumptions are abundant in the literature, showing the need and importance of this consistency test.

Using two methods to combine measurements of 2D BAO and SNe data, we estimated values of the 
absolute BAO scale ranging from $141.45\; {\rm{Mpc}} \leq r_s \leq 159.44\; {\rm{Mpc}}$ (Binning) and $143.35 \; {\rm{Mpc}} \leq r_s \leq 161.59\; {\rm{Mpc}}$ (GP) for eight different measurements of $H_{0}$. The results from both methods agree with each other and with the CMB estimate of the sound horizon at drag epoch, $r_{d} = 147.09 \pm 0.26$ Mpc at 1$\sigma$ \cite{Planck2018}, demonstrating the robustness of the $\Lambda$CDM model. However, it is important to emphasize that such compatibility is found because our model-independent approach provides $\sim 10\%$-error estimates on $r_{d}$, which comes mainly from the current uncertainties of $\theta(z)$ measurements. Therefore, our analysis and results show the potential of the consistency test discussed in this paper, attest to the robustness of the $\Lambda$CDM model at high-$z$ from the current data, and also reinforce the need for more precise measurements of the 2D BAO scale, which is expected from the upcoming data of the new generation of galaxy surveys \cite{J-PAS:2014hgg,Bonoli:2020ciz,DESI:2016fyo,EuclidTheoryWorkingGroup:2012gxx}.

\section*{Acknowledgements}

TL thanks the financial support from the Coordena\c{c}\~ao de Aperfei\c{c}oamento de Pessoal de N\'{\i}vel Superior (CAPES). Ruchika acknowledges the funding from IIT Bombay, where part of the work has been done. JSA is supported by Conselho Nacional de Desenvolvimento Cient\'{\i}fico e Tecnol\'ogico (grant no. 307683/2022-2) and Funda\c{c}\~ao de Amparo \`a Pesquisa do Estado do Rio de Janeiro (FAPERJ) grant 259610 (2021).

\appendix{}
\section{Supplementary Figures}\label{Appendix}
\begin{figure*}
\begin{center}
\resizebox{200pt}{130pt}{\includegraphics{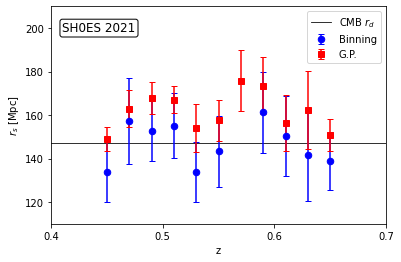}}
\hspace{1mm}\resizebox{200pt}{130pt}{\includegraphics{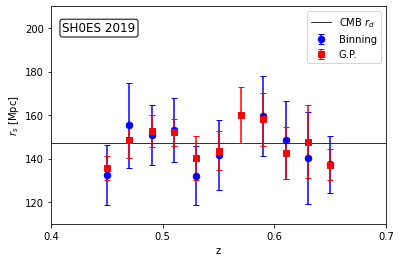}}\\
\hspace{1mm}\resizebox{200pt}{130pt}{\includegraphics{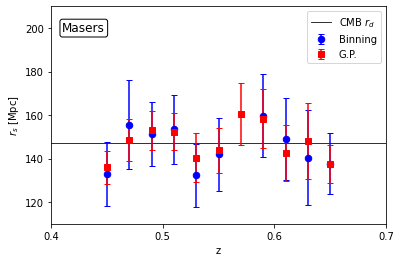}}
\resizebox{200pt}{130pt}{\includegraphics{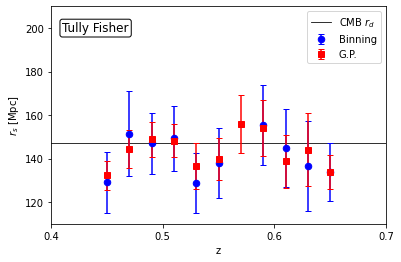}} 
\resizebox{200pt}{130pt}{\includegraphics{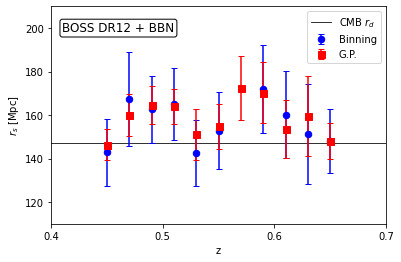}}
\hspace{1mm}\resizebox{200pt}{130pt}{\includegraphics{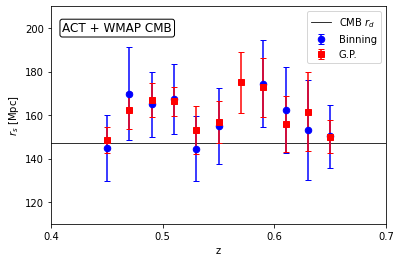}}
\resizebox{200pt}{130pt}{\includegraphics{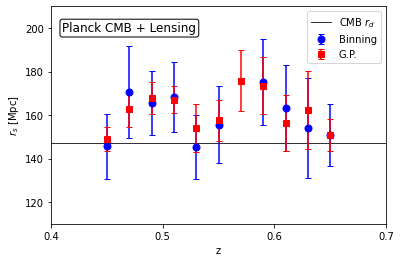}}\\
\end{center}
\caption{The acoustic scale of BAO from transversal (2D) BAO and SNe datasets assuming different measurements and estimates of the Hubble constant. The horizontal line represents the current estimate of the sound horizon at drag epoch from CMB observations assuming the $\Lambda$CDM model, $r_{d} = 147.09 \pm 0.26$ Mpc (1$\sigma$) \cite{Planck2018}.}  
\label{lowexp}
\end{figure*}

\end{document}